\documentstyle[preprint,12pt,aps]{revtex}
\tightenlines
\newif\ifboo \boofalse
\def\And{{\rm and\ }}
\def\Review#1{\boofalse{\it #1},}
\def\Name#1{{\sc #1},}
\def\Vol#1{\ifboo Vol. {\bf #1}\else{\bf #1}\fi}
\def\Year#1{\ifboo #1\else(#1)\fi}
\def\Book#1{\bootrue{\it #1},}
\def\Page#1{\ifboo {\rm p. #1}\else{\rm #1}\fi}
\begin{document}
%
\title{Triplet interactions in star polymer solutions}
\author{C. von Ferber, A. Jusufi, C. N. Likos, H. L\"owen, M. Watzlawek}
\address{Institut f\"ur Theoretische Physik II, Heinrich-Heine-Universit\"at 
D\"usseldorf, D-40225 D\"usseldorf, Germany
}
\maketitle
\date{\today}
\begin{abstract}
  We analyze the effective triplet interactions between the centers of
  star polymers in a good solvent. Using an analytical short distance
  expansion inspired by scaling theory, we deduce that the triplet
  part of the three-star force is attractive but only 11\% of the
  pairwise part even for a close approach of three star polymers.  We
  have also performed extensive computer simulations for different arm
  numbers $f$ to extract the effective triplet force.  The simulation
  data show good correspondence with the theoretical predictions. Our
  results justify the effective pair potential picture even beyond the
  star polymer overlap concentration.
  \\
  PACS {82.70.Dd, 64.60.Fr, 61.20.Ja, 61.41.+e}
\end{abstract}
\section{Introduction}
Star polymers \cite{Grest96}, i.e., structures of $f$ linear polymer
chains that are chemically linked with one end to a common core, have
found recent interest as very soft colloidal particles
\cite{Gast96,Seghrouchni98,Likos98,Watzlawek99,Jusufi99}.  As the
number $f$ of chains increases, they interpolate between linear
polymers and polymeric micelles \cite{Grest96,Gast96,Fleischer99}.
For large $f$, the effective repulsion between the cores of different
polymer stars becomes strong enough to allow for crystalline ordering
in a concentrated star polymer solution. While such a behavior was
already predicted by early scaling arguments
\cite{Witten86a,Witten86b} only recently corresponding experiments
have become feasible with sufficiently dense star solutions. The
crystallization transition occurs roughly at the overlap concentration
$c^*$ which is the number density of stars where their coronae start
to touch experiencing the mutual repulsion. It is defined as
$c^{*}=1/(2R_g)^3$ where $R_g$, the radius of gyration, is the root mean 
square distance of the monomers from the center of mass of a single star.  
In addition, theory and computer simulation have refined the
original estimate for the number of chains $f$ necessary for a
freezing transition from $f\sim 100$ \cite{Witten86a,Witten86b} to
$f\sim 34$ \cite{Watzlawek99} and predicted a rich phase diagram
including stable anisotropic and diamond solid structures at high
densities and high arm numbers.  These results were derived using an
effective {\it pair potential} between stars with a logarithmic short
distance behavior derived from scaling theory.

In general, while the pair interactions are the central focus and the
typical input of any many-body theory, much less is known about
triplet and higher-order many body interactions. For rare gases, the
Axilrod-Teller triplet interaction \cite{Axilrod1943} has been found
to become relevant in order to describe high-precision measurements of
the structure factor \cite{Tau1989}. For charged colloids, the
effective triplet forces are generated by nonlinear counterion
screening. This was investigated recently by theory and simulations
\cite{Loewen1998}. For star polymer solutions in a good solvent such
studies are missing.  In all three cases, the effective triplet forces
originate from formally integrating out microscopic degrees of
freedom. For rare gases, these are the fluctuations of the outer-shell
electrons while for charged colloids the classical counterions play
the role of additional microscopic degrees of freedom.  For star
polymers, on the other hand, one is interested in an effective
interaction between the star centers by integrating out the monomer
degrees of freedom \cite{Graf98}.  Usually one starts from an
effective pair potential which is valid for large particle separation.
The range of this effective pair potential involves a certain length
scale $\ell$ which is the decay length of the van-der-Waals
attraction, the Debye-H\"uckel screening length or the diameter of
gyration $2R_g$, for rare gases, charged colloids, and star polymers,
respectively. Triplet forces, i.e. three star forces, not forces
between monomers, become relevant with respect to the pairwise forces
if the typical separations between the particles are smaller than this
typical length scale $\ell$. This implies a triple overlap of particle
coronae drawn as spheres of diameter $\ell$ around the particle
centers. The triple overlap volume is an estimate for the magnitude of
the triplet forces. Hence a three-particle configuration on an
equilateral triangle is the configuration where triplet effects should
be most pronounced.

The aim of the present paper is to quantify the influence of triplet
interactions for star polymer solutions in a good solvent using both
analytical theory and computer simulation.  In doing so, we consider a
set-up of three star polymers whose centers are on an equilateral
triangle.  We found that the triplet part is attractive but its
relative contribution is small (11\%) with respect to the repulsive
pairwise part. This relative correction is universal, i.e., it is
independent of the particle separation and of the arm number.
It even persists for a collinear configuration of three star polymers
where the absolute correction is smaller than in the triangular 
situation for the same star-star distance.
Consequently, the validity of the effective pair potential model is
justified even at densities above the overlap concentration.  In
particular, our result gives evidence that the anisotropic and diamond
solids predicted by the pair theory are indeed realizable in actual
samples of concentrated star polymer solutions.

Our paper is organized as follows: in section II we apply scaling
theory to extract the triplet forces both for small and for large arm
numbers.  In section III we briefly describe our Molecular Dynamics
(MD) simulation scheme and present results in section IV.  Comparing
these to the theoretical predictions, we find good agreement. Section
V is devoted to concluding remarks and to an outlook.

\section{Scaling theory of triplet forces between  star polymers}
\subsection{Scaling of single stars}
The scaling theory of polymers was significantly advanced by de
Gennes' observation that the $n$-component spin model of magnetic
systems is applicable to polymers in the formal $n=0$ limit
\cite{deGennes72}.  This opened the way to apply renormalization group
(RG) theory to explain the scaling properties of polymer solutions
that have been the subject of experimental and theoretical
investigations since the pioneering works in this field
\cite{Flory69}.  Many details of the behavior of polymer solutions may
be derived using the RG analysis \cite{Schaefer99}. Here, we use only
the more basic results of power law scaling: the radius of gyration
$R_g(N)$ of a polymer chain and the partition function ${\cal Z}(N)$
are found to obey the power laws:
\begin{equation}\label{1}
R_g(N)\sim N^\nu \mbox{\hspace{3em} and \hspace{3em}}
{\cal Z}(N)\sim z^N N^{\gamma-1} \,.
\end{equation}
The fugacity $z$ measures the mean number of possibilities to add one
monomer to the chain. It is microscopic in nature and will depend on
the details of the model or experimental system.  The two exponents
$\nu$ and $\gamma$ on the contrary are the $n=0$ limits of the
correlation length exponent $\nu(n)$ and the susceptibility exponent
$\gamma(n)$ of the $n$ component model and are universal to all
polymer systems in a good solvent, i.e., excluding high concentration
of polymers or systems in which the polymers are collapsed or are near
the collapse transition.  For any such system the exponents of any
other power law for linear polymers may be expressed by these two
exponents in terms of scaling relations.

It has been shown that the $n$ component spin model may be extended by
insertions of so called composite spin operators that allow to
describe polymer networks and in particular star polymers in the $n=0$
limit \cite{Duplantier86,Ohno88,Schaefer92}.  A family of additional
exponents $\gamma_f$ governs the scaling of the partition function
${\cal Z}_f(N)$ of a polymer star of $f$ chains each with $N$
monomers:
\begin{equation}\label{2}
{\cal Z}_f(N)\sim z^N N^{\gamma_f-1} .
\end{equation}
Again the exponents of any other power law for more general polymer
networks are given by scaling relations in terms of $\gamma_f$ and
$\nu$.  Here, we substitute another family of exponents $\eta_f$ to
replace $\gamma_f-1=\nu(\eta_f-f\eta_2)$. The first two members
$\eta_1=0$ and $\eta_2=(1-\gamma)/\nu$ are defined by the requirement
that the $f=1$ star and the $f=2$ star are just linear chains with
scaling exponents $\gamma_1=\gamma_2=\gamma$.  The values of these
exponents are known from renormalization group analysis (RG) and Monte
Carlo (MC) simulations \cite{Batoulis89}.  Several equivalent
approaches have been elaborated to evaluate the renormalized
perturbation theory.  Early first order perturbative RG results were
given in ref. \cite{Miyake83}.  Here, we explicitly present the result
of an expansion in the parameter $\varepsilon=4-d$ where $d$ is the
space dimension.  The $\varepsilon$-expansion for the $\eta_f$ reads
\cite{Schaefer92}
\begin{eqnarray}
\eta_f = -\frac{\varepsilon}{8}f(f-1)\Big\{ 1-\frac{\varepsilon}{32}(8f-25)
 +\frac{\varepsilon^2}{64}\Big[(28f-89)\zeta(3)+8f^2-49f+\frac{577}{8}\Big]
\Big\} +{\cal O}(\varepsilon^4)
 \label{2a}
\end{eqnarray}
with the Riemann $\zeta$-function.  Note that this series is
asymptotic in nature and to evaluate it for $\varepsilon=1$ it is
necessary to apply resummation. An alternative expansion for the star
exponents makes use of an RG approach at fixed dimension $d=3$
proposed by Parisi \cite{Parisi80}.  This expansion has been worked
out in refs. \cite{FerHol96a,FerHol97c,FerHol97d}.  The corresponding
expressions are lengthy and not presented here.  In Table 1, in the
first two lines we have calculated the resummation for the series in
Eq. (\ref{2a}) as well as for the expansion at fixed dimension.  The
resummation procedure that we apply combines a Borel transform with a
conformal mapping using all information on the asymptotic behavior of
the perturbation expansion of the corresponding spin model
\cite{Brezin77,LeGuillou80}.  Results for $f\leq 9$ have been given
before in refs.  \cite{Schaefer92,FerHol96a,FerHol97c,FerHol97d}
whereas we have added here the calculation of values for $f=10,12,15$.
The deviation between the two approaches measures the error of the
method.  For large $f$ the leading coefficient of the 
$k$th order term $\varepsilon^k$ in Eq. (\ref{2a}) is
multiplied by $f^{k+1}$.  This is due to combinatorial reasons and
occurs also for the alternative approach. It limits the use of the
series to low values of $f$.

Another possibility to estimate the values of the star polymer scaling
exponents $\gamma_f$ is to consider the limiting case of many arm star
polymers.  For large $f$ each chain of the star is restricted
approximately to a cone of solid angle $\Omega_f=4\pi/f$. In this cone
approximation one finds for large $f$ \cite{Ohno89}
\begin{equation}
\gamma_f\sim -f^{3/2}.
\end{equation}
\subsection{Two star polymers}
Let us now turn to the effective interaction between the cores of two
star polymers at small distances $r$ that are small on the scale of
the size $R_g$ of the stars.  Let us for the moment consider a more
general case of two star polymers with $f_1$ and $f_2$ arms
respectively. The cores of the two stars are at a distance $r$ from
each other.  We assume all chains involved to be of the same length.
The power law for the partition sum ${\cal Z}^{(2)}_{f_1f_2}(r)$ of
two star polymers may then be derived from a short distance expansion.
This expansion is originally established in the field theoretic
formulation of the $n$ component spin model. While we do not intend to
give any details of these considerations here, applications to polymer
theory may be found in refs. \cite{Duplantier89,Ferber97}. The
relevant result on the other hand is simple enough: the partition sum
of the two stars ${\cal Z}^{(2)}_{f_1f_2}(N,r)$ at small distance $r$
factorizes into a function $C_{f_1f_2}(r)$ of $r$ alone and the
partition function ${\cal Z}_{f_1+f_2}$(N) of the star with $f_1+f_2$
arms that is formed when the cores of the two stars coincide.
\begin{equation}\label{4}
{\cal Z}^{(2)}_{f_1f_2}(N,r)\sim C_{f_1f_2}(r) {\cal Z}_{f_1+f_2}(N)
\end{equation}
For the function $C_{f_1f_2}(r)$ one may show that power law scaling for 
small $r$ holds in the form
\begin{equation}
C_{f_1f_2}(r)\sim r^{\Theta^{(2)}_{f_1f_2}}\,.
\end{equation}
with the contact exponent $\Theta^{(2)}_{f_1f_2}$.  To find the
scaling relation for this power law we we change the length scale in
(\ref{4}) in an invariant way by $r\to \lambda r$ and $N\to
\lambda^{1/\nu}N$.  The scaling of the partition function ${\cal
  Z}^{(2)}_{f_1f_2}$ may be shown to factorize into the contributions
for the two stars. This transforms (\ref{4}) to
\begin{equation}
 \lambda^{-1/\nu(\gamma_{f_1}-1)} \lambda^{-1/\nu(\gamma_{f_2}-1)}
{\cal Z}^{(2)}_{f_1f_2}(\lambda^{1/\nu}N,\lambda r) \sim
\lambda^{-\Theta^{(2)}_{f_1f_2}}C_{f_1f_2}( \lambda r)
\lambda^{-1/\nu(\gamma_{f_1+f_2}-1)}{\cal Z}_{f_1+f_2}(\lambda^{1/\nu}N)\,.
\end{equation}
Collecting powers of $\lambda$ provides the scaling relation
\begin{eqnarray}
\nu\Theta^{(2)}_{f_1f_2} &=& (\gamma_{f_1}-1)+(\gamma_{f_2}-1) 
- (\gamma_{f_1+f_2}-1)\,,
\nonumber\\
\Theta^{(2)}_{f_1f_2} &=&  \eta_{f_1}+\eta_{f_2}-\eta_{f_1+f_2}\,.
\end{eqnarray}

We now specialize our consideration to the interaction between two
stars of equal number of arms $f_1=f_2=f$.  The mean force
$F^{(2)}_{ff}(r)$ between the two star polymers at short distance $r$
is then easily derived from the effective potential 
$V^{\rm eff}(r)=
-k_{\rm B}T\log[ {\cal Z}^{(2)}_{ff}(r)/({\cal Z}_{f})^{2}]$
with $k_{\rm B}T$ denoting the thermal energy. For the force this
results in
\begin{equation}\label{7}
 \frac{1}{k_{\rm B}T}F^{(2)}_{ff}(r)= 
  \frac{\Theta^{(2)}_{ff}}{r}\,.
\end{equation}
The cone approximation for the contact exponents \cite{desCloizeaux75}
$\Theta^{(2)}_{ff}$ may be matched to the
known values for $f=1,2$
(see table 1), fixing the otherwise unknown prefactor. Assuming that
the behavior of the $\Theta^{(2)}_{ff}$ may be described by the cone
approximation for all $f$ one finds:
\begin{equation}
 F^{(2)}_{ff}(r)\approx\frac{5}{18} \frac{f^{3/2}}{r} \,.
\end{equation}
This matching in turn suggests an approximate value for the $\eta_f$ 
exponents,
\begin{equation}\label{8}
 \eta_f\approx -\frac{5}{18} (2^{3/2}-2)^{-1}f^{3/2} \,.
\end{equation}
Note on the other hand that this approximation is inconsistent with
the exact result $\eta_1=0$. However, the approximation works well for
$\Theta^{(2)}_{ff}$ in the range $f=1,\ldots,6$ were we have
calculated the corresponding values from the perturbation theory
results as well as according to the cone approximation.  Our results,
displayed in the second part of table 1, show good correspondence of
the cone approximation with the resummation values.
     
\subsection{Three stars}
We now use the idea of the short distance expansion once more to
derive the triplet interaction of three star polymers at close
distance.  We consider a symmetric situation in which the three cores
of the polymer stars are located on the corners of an equilateral
triangle (see Fig. 1).  The distance between the cores is $r$ while
their distance to the center of the triangle is $R$. We assume that
the radius of gyration $R_g$ of the star polymers is much larger than
their mutual distance $R_g\gg r$.

To make the argument more transparent we first consider the slightly
more general case of three stars with $f_1$, $f_2$ and $f_3$ arms
respectively.  Shrinking the outer radius $R$ of the triangle on which
the cores are located, the partition function of this configuration of
three stars will scale with $R$ according to
\begin{eqnarray}
{\cal Z}_{f_1f_2f_3}(R)&\sim& R^{\Theta^{(3)}_{f_1f_2f_3}}\\
\Theta^{(3)}_{f_1f_2f_3}&=& \eta_{f_1}
+\eta_{f_2}+\eta_{f_3}-\eta_{f_1+f_2+f_3}\,.
\end{eqnarray}
Now, the scaling exponent $\eta_{f_1+f_2+f_3}$ of the star that
results by collapsing the cores of the three stars at one point has to
be taken into account as follows from an argument analogous to the
above consideration for two stars.

Let us specify the result for the symmetric situation of three
equivalent stars $f_1=f_2=f_3=f$. Furthermore we assume that the large
$f$ approximation (\ref{8}) is valid for the exponents $\eta_f$. Then
the three star contact exponent may be written as
\begin{equation}
\Theta^{(3)}_{fff}=\frac{3^{3/2}-3}{2^{3/2}-2}\times\frac{5}{18}f^{3/2}\,.
\end{equation}
An effective potential of the system of the three stars at small
distance $R$ from the center may then be defined by
\begin{equation}
V^{(3)\rm eff}_{fff}(R) = -k_{\rm B}T\Theta^{(3)}_{fff} \ln(R/R_g)\,.
\end{equation}
We now derive the corresponding three body force underlying this
effective potential. Note that the absolute value of the force is the
same for all three stars.  The relation of the potential to the force
on the core of one star is then
\begin{equation}
V^{(3)\rm eff}_{fff}(R+dR)-V^{(3)\rm eff}_{fff}(R)=
\sum_{i=1}^3\vec{F_i}\cdot d\vec{R}_i=3F_{fff}^{(3)}(R)dR\,.
\end{equation}
The final result for the total force on each of the stars that
includes any three body forces is therefore
\begin{equation}
F^{(3)}_{fff}(R)=-k_{\rm B}T\Theta^{(3)}_{fff}/(3R)\,.
\end{equation}
If one starts instead from a sum of two body forces, then one star
experiences the sum of the two forces calculated for the star-star
interaction. With the given geometry of the equilateral triangle this
is easily calculated to give
\begin{equation}\label{14}
F^{(2)}_{fff}(r)=|\hat{r}_{12}\Theta^{(2)}_{ff}/r_{12}+
\hat{r}_{13}\Theta^{(2)}_{ff}/r_{13}|=-k_{\rm B}T\Theta^{(2)}_{ff}/R\,.
\end{equation}
Here, $r=r_{12}=r_{13}=R\sqrt{3}$ denote the distance between two of
the stars, while the $\hat{r}_{ij}$ are the unit vectors along the
edges of the triangle (see Fig.1).  The relative deviation from the
pair potential picture is then given by
\begin{equation}\label{15}
\frac{\Delta F}{F^{(2)}_{fff}}=
\frac{F^{(3)}_{fff}(r)-F^{(2)}_{fff}(r)}{F^{(2)}_{fff}(r)}
=\frac{\Theta^{(3)}_{fff}-3\Theta^{(2)}_{ff}}{3\Theta^{(2)}_{ff}}\,.
\end{equation}
Using the cone approximation for the contact exponent we finally
obtain for the relative deviation caused by triplet forces alone
\begin{equation}\label{15a}
\frac{\Delta F}{F^{(2)}_{fff}}=\frac{3^{3/2}-3}{2^{3/2}-2}\approx -0.11\,.
\end{equation}
This result is independent of the number of arms and valid in the full
region that is described by the logarithmic potential. In table 1 we
have calculated the exponents as derived from the perturbation
expansion of polymer field theory \cite{Ferber97,FerHol97c,FerHol97d}
checking the relation eq. (\ref{15a}).  Taking into account the error
that may be estimated from the difference of the results obtained by
the two complementary approaches, the results are in good agreement
with the cone approximation even for low $f$ values.  The fair
coincidence is rather surprising as additional numerical errors might
be introduced by the calculation of the contact exponents from the
original star exponents. It confirms our estimate of the relative
deviation caused by triplet forces to be of the order of not more than
$11\%$ for all analytic approaches we have followed here.
Let us note that the analogous calculation for a symmetric linear 
configuration of three stars yields the same relative deviation eq. 
(\ref{15a}). The absolute triplet forces for the linear configuration 
are smaller by a factor $\sqrt{3}/2$ than for the triangular configuration
with the same star-star distance.  
\section{Computer simulation method}
Molecular dynamics (MD) simulations were performed using exactly the
model that three of the present authors devised to test the effective
pair potential \cite{Jusufi99} and had been originally proposed to
study single star polymers \cite{Grest87,Grest94}.  In this model the
configuration of star polymer $i=1,2,3$ is given by the coordinates
${\vec r}^{(i,j)}_m$ of the $N$ monomers $m=1,\ldots,N$ of the $f$
chains $j=1,\ldots,f$ and the position of its core $r^{(i)}_0$.  The
main features of this model are the following: (1) A purely repulsive
truncated Lennard-Jones like potential acts between all monomers
$m=0,\ldots,N$ on all chains.  (2) An attractive FENE-potential
\cite{Grest87,Grest94} that preserves the chain connectivity and acts
only between consecutive monomers $m,m+1$ along each chain.  (3) These
potentials have to be slightly modified for the interaction between
the first monomer $m=1$ and the core $m=0$ of the star to allow the
core to have a radius $R_d$ that is sufficiently large to place $f$
monomers in its vicinity.

The three cores of the stars were placed at the corners of an
equilateral triangle, see again Figure 1 where also the core radius
$R_d $ is shown.  A typical snapshot of the three star simulation is
displayed in Figure 2 for a functionality of $f=5$ and $N=100$
monomers per chain.  The force on the star core was averaged during
the MD simulation for a number of edge lengths $r$ of the triangle
varying in the range between the diameter of the two cores $2R_d$ and
the diameter of gyration $2R_g$ of a single star polymer. We have
produced data for $f=3,5,10,18,30$.  For the smaller functionalities
($f=3,5,10$) the number of monomers per chain was $N=100$ while for
$f=10,18,30$ a number $N=50$ was chosen. Note that the total system
comprises between $900-4500$ mutually interacting particles. As
equilibration is slow and the statistical average converges slowly,
the simulation becomes increasingly time-consuming beyond such system
sizes.  As for reference data, we have also produced data for a two
stars situation according to the calculations in Ref.\ 
\cite{Jusufi99}.

\section{Results}
Results of the computer simulation are compared to the theory in
Figures 3a and 3b.  The reduced averaged force on a single star is
shown versus the reduced triangle length for different arm numbers. As
a reference case, also the corresponding results in a pair potential
picture are shown, both within theory and simulation.  For technical
reasons we kept a small core radius $R_d$ in the simulation, which is
roughly $10\%$ of the radius of gyration of the whole star.  In the
theory, on the other hand, the core size was zero. Hence, to compare
properly \cite{Jusufi99}, a shift $r-2R_d$ has to be performed.

As expected, in both theory and simulation, the triplet forces become
relevant only within the coronae. A comparison with pure pairwise
forces leads to the first important observation that the triplet force
is smaller, i.e.\ the pure triplet contribution is {\it attractive}.
(Note that one has to multiply the pure two-star force by a factor of
$\sqrt{3}$ for simple geometrical reasons.)  The relative magnitude of
the triplet term, however, is small.  A quantitative comparison with
theory and simulation leads to good overall agreement. The triplet
contribution itself, however, is subjected to larger statistical
errors of the simulation. Hence we resorted to a different strategy to
check the theory by plotting the inverse force versus distance.  If
the theory is correct the simulation data should fall on a straight
line both for the pure pairwise and the full triplet case. The slope
should then give the theoretical prefactor of the logarithmic
potential.  The advantage of this consideration is that the slope
bears a smaller statistical error as more data points are included.
Such a comparison is shown in Figure 4 for $f=10$.  The first
consequence is that the simulation data indeed fall on a straight line
confirming the theory. In fact this is true for all other parameter
combinations considered in the simulations.  The slope is higher for
the triplet and lower for the pair case, both in theory and
simulation. The actual values in Figure 4 are in the same order of
magnitude but a bit different.

In order to check this in more detail, we have extracted the slope for
all simulation data. The result is summarized in Figure 5 where the
relative differences of the slopes between the pair and triplet cases
are plotted versus the arm number $f$.  The theory predicts a constant
value of $0.11$, see Eq. (\ref{14}).  The simulation data scatter a
lot in the range between $0.05$ and $0.15$ due to the large
statistical error but the theoretical value falls reasonably within
the data.  Consequently, the triplet contributions are found to be
attractive and small even for nearly touching cores where the triplet
overlap of the coronae is substantial.

\section{Conclusions}

In conclusion, we have calculated, by theory and computer simulations,
the triplet interaction between star polymer centers in a good solvent
positioned on the corners of an equilateral triangle. The triplet part
was found to be attractive but only about $11\%$ of the pairwise
repulsion. Our calculations justify earlier investigations
\cite{Watzlawek99} where the pair potential framework was used even
slightly above the star overlap concentration.

We finish with a couple of remarks: First, the scaling theory can also
be performed for any triplet configurations beyond the
equilateral triangle studied in this paper. Second, arbitrary
higher-order many body forces can be investigated assuming a cluster
of $M$ stars. Such a calculation is given in Appendix A.  As a result,
the deviations from the pair potential picture increase with the
number $M$ and even diverge for $M\to \infty$.  This implies that the
pair potential picture breaks down for very high concentrations.  This
is expected as for high concentration a star polymer solution is
mainly a semi-dilute solution of linear chains where it is irrelevant
at which center they are attached to \cite{Witten86a}. As far as
further simulational work is concerned, there are many open problems
left.  Apart from the investigation for arbitrary triplet
configurations and their extensions to an arbitrary number of stars,
the most challenging problem is a full ``ab initio" simulation of many
stars including many-body forces from the very beginning. This is in
analogy to Car-Parrinello simulations \cite{Car85} which were also
applied to colloidal suspensions \cite{Loewen1992}. A first attempt
has been done \cite{Pakula98}, but certainly more work is needed here.
Another (a bit less demanding) task is to study stars on a periodic
solid lattice with periodic boundary conditions and extract the many
body interactions from there.

It would be interesting to study the relevance of triplet forces for
star polymers in a {\it poor} solvent near the $\Theta$-point
\cite{Likos98a}. It can, however, be expected that the triplet forces
here are even less important than for a good solvent as the effective
interaction becomes stiffer in a poor solvent. Furthermore, the effect
of polydispersity in the arm number which has been briefly touched in
our scaling theory treatment should be extended since this is
important to describe real experimental samples.
 
\section*{Acknowledgments}

We are grateful to the DFG for financial support within the SFB 237.

\begin{appendix}
\renewcommand{\thesection}{\Alph{section}}%
\renewcommand{\theequation}{\Alph{section}.\arabic{equation}}%
\section{Higher Order forces between star polymers }\label{A} 
Here, we derive for the general case of $M$ simultaneously interacting
star polymers with $f$ arms the effective $M$th order force.
Generalizing the equilateral triangle geometry, we study the situation
where the $M$ cores of the stars are evenly distributed on a sphere
with radius $R$. In particular, the cores of the stars may be located
at the corners of a regular polyhedron. Then the non-radial forces on
each star polymer cancel.  The latter condition may be fulfilled
approximately also for large numbers $M$ for which a regular
polyhedron does not exist.

We first calculate the force on one star by the sum of $M-1$ pairwise
forces effected by the other stars.  For the pairwise force (\ref{7})
that acts according to a $1/r$-law it is easy to verify that the
radial component of the force between any two points on the sphere is
$\Theta^{(2)}_{ff}/(2R)$ independent of their relative position. With
this simplification the total (radial) force on one star is
\begin{equation}
\frac{1}{k_{\rm B}T}F^{(2)}_{M,f}=\frac{M-1}{2} 
\frac{\Theta^{(2)}_{ff}}{R}\,.
\end{equation}
Here, $F^{(2)}_{M,f}$ denotes the sum of pairwise forces on one of the
$M$ stars each with $f$ arms.  In the case $M=3$ this is the result of
eq. (\ref{14}).

The total $M$th order force $F^{(M)}_{M,f}$ between $M$ star polymers
with $f$ arms brought close together may again be derived from a short
distance expansion resulting in the scaling relation
\begin{equation}
\Theta^{(M)}_{M,f}=M\cdot\eta_f-\eta_{M\cdot f}\,.
\end{equation}
The force on one star is then found in the same way as for three stars
as
\begin{equation}
\frac{1}{k_{\rm B}T}F^{(M)}_{M,f}=\frac{\Theta^{(M)}_{M,f}}{M\cdot R}\,.
\end{equation}
The leading contributions for large numbers of stars $M$ in the 
two cases differ even in the power of $M$. 
While the first is linear in $M$ the latter grows only with the square root
of $M$.  In the large-$f$ and large-$M$ approximations
this reads : 
\begin{eqnarray}
\frac{1}{k_{\rm B}T}F^{(2)}_{M,f}&\approx& 
\frac{5}{18}\frac{f^{3/2}}{2R}M \\
\frac{1}{k_{\rm B}T}F^{(M)}_{M,f}&\approx& 
\frac{5}{18}(2^{3/2}-2)\frac{f^{3/2}}{R}M^{1/2}\,.
\end{eqnarray}
Note that for large $M$ the factors $M$ and $M^{1/2}$ in these two
approaches are not a result of the large $f$ approximation but are of
combinatorial and geometrical origin.  This shows that for large $M$
the sum of pairwise forces largely overestimates the force on one
star.
\end{appendix}
\newpage
\begin{table}
\begin{tabular}{lrrrrrrrrrrr}
f         &1&   2 &   3 &   4 &  5 &   6 &   8 &   9 & 10 &  12 & 15 \\
\hline \\
a $\eta_f$&0&-0.28&-0.75&-1.36&-2.07&-2.88&-4.71&-5.72&-6.80&-9.12&-12.98\\
b         &0&-0.28&-0.76&-1.38&-2.14&-3.01&-5.06&-6.22&-7.48&-10.23&-14.93\\
\\
a $\Theta^{(2)}_{ff}$
       &0.28& 0.80& 1.38& 1.99& 2.66& 3.36 \\
b      &0.28& 0.82& 1.49& 2.30& 3.20& 4.21 \\
c      &0.28& 0.79& 1.44& 2.22& 3.11& 4.08 \\
\\
a $\Theta^{(3)}_{fff}$
       &0.75& 2.04& 3.47& 5.04& 6.77 \\
b      &0.76& 2.17& 3.94& 6.09& 8.51 \\
c      &0.74& 2.08& 3.83& 5.89& 10.82 \\
\\
a $\Delta F / F$  
       &-0.11&-0.15&-0.16&-0.16&-0.15  \\
b      &-0.09&-0.12&-0.12&-0.12&-0.11  \\
c      &-0.11&-0.11&-0.11&-0.11&-0.11
\end{tabular}
\caption{
Calculation of the exponents that govern the pair and triplet interactions. 
The labels (a) and (b) stand for the two complementary renormalization
group approaches (expansion in $\varepsilon=4-d$ and massive renormalization 
at $d=3$) used to calculate the exponents $\eta_f$.
The difference
of the two results may be taken as an estimation of the error of the method.
Label (c) stands for the cone approximation result with matching to $f=1,2$
as explained in the text.}
\end{table}


\newpage
\section*{Figure Captions}
FIG. 1.  Three star polymers at mutual distance $r$. The cores of the
stars (with radius $R_d$) are located at the corners of an equilateral
triangle. The distance from the center is $R$.  The mean radius of
gyration of a single star is $R_g$.

\vspace{3ex}

FIG. 2.  Snapshot of the simulation of three stars with $f=5$ arms
each with $N=100$ monomers. The cores are located at the corners of
the equilateral triangle that is depicted in the center. The monomers
that belong to the same star are represented by balls of the same
color: either black, dark gray, or light gray.

\vspace{3ex}

FIG. 3a.  Comparison of the force $F$ measured in the three star MD
with that calculated from a corresponding two star MD simulation for
$f=3$ and $f=10$ with $N=100$. Also the results predicted by the
theory are plotted as a continuous line (only pair forces) and a
broken line (including triplet forces).

\vspace{3ex}

FIG. 3b.  Same as Fig. 3a but for $f=18$ and $f=30$ with $N=50$.

\vspace{3ex}

FIG. 4.  Comparison of the inverse force $1/F$ measured in the three
star MD with that calculated from a corresponding two star MD
simulation for $f=10$ with $N=50$.  The linear fits for the pair
forces (small dashed line) and the full three body force (dash-dotted
line) are shown together with the respective results predicted by the
theory which are depicted by a continuous line (only pair forces) and
a broken line (including triplet forces).

\vspace{3ex}

FIG. 5.  The slopes of the linear fits to the data as shown in Fig. 4
were extracted from the simulation data for $f=3,5,10,18,30$ and
$N=50,100$ to calculate the relative deviation $\Delta
F/F^{(2)}_{fff}$ induced by the triplet forces. The line at $0.11$
corresponds to the analytic result.

\end{document}